\documentclass[%
 reprint,
 superscriptaddress,
 longbibliography,
 amsmath,amssymb,
 aps,
 prx,
floatfix,
]{revtex4-2}
\usepackage[utf8]{inputenc}
\usepackage[english]{babel}
\usepackage{graphicx}
\usepackage{dcolumn}
\usepackage{bm}
\usepackage{comment}
\usepackage{xcolor}
\usepackage{makecell}
\usepackage{booktabs}
\usepackage{amsmath}
\usepackage{braket}
\usepackage[normalem]{ulem}
\usepackage{MnSymbol}
\usepackage{natbib}
\usepackage{chngcntr}
\counterwithout{figure}{section}
\usepackage{todonotes}
\usepackage{subfiles} 

\usepackage[autostyle]{csquotes}
\usepackage[colorlinks,citecolor=red,urlcolor=blue,bookmarks=false,hypertexnames=true]{hyperref}

\newcommand{\rev}[1]{\textcolor{black}{#1}}

\begin{document}

\makeatletter
\renewcommand{\p@subsection}{}
\renewcommand{\p@subsubsection}{}
\makeatother

\preprint{APS/123-QED}
\title{Inverse-Designed Photonic Crystal Cavities with Controllable Far-Field Numerical Aperture}

\author{Neelesh Kumar Vij *}
\affiliation{Institute for Research in Electronics and Applied Physics and Joint Quantum Institute, University of Maryland, College Park, Maryland 20742, USA}
\thanks{\textit{Corresponding author: Neelesh Kumar Vij (nkvij@umd.edu)}}
\affiliation{Department of Electrical and Computer Engineering, University of Maryland, College Park, Maryland 20740, USA}
\author{Purbita Purkayastha}
\affiliation{Institute for Research in Electronics and Applied Physics and Joint Quantum Institute, University of Maryland, College Park, Maryland 20742, USA}
\affiliation{Department of Physics, University of Maryland, College Park, Maryland 20740, USA}
\author{Jasvith Raj Basani}
\affiliation{Institute for Research in Electronics and Applied Physics and Joint Quantum Institute, University of Maryland, College Park, Maryland 20742, USA}
\affiliation{Department of Electrical and Computer Engineering, University of Maryland, College Park, Maryland 20740, USA}
\author{Edo Waks}
\affiliation{Institute for Research in Electronics and Applied Physics and Joint Quantum Institute, University of Maryland, College Park, Maryland 20742, USA}
\affiliation{Department of Electrical and Computer Engineering, University of Maryland, College Park, Maryland 20740, USA}
\affiliation{Department of Physics, University of Maryland, College Park, Maryland 20740, USA}

\begin{abstract}
Photonic crystal cavities confine light to subwavelength volumes, enabling strong light-matter interactions for applications in low-power photonics, optoelectronics, nonlinear optics, and quantum information. These applications demand cavities that combine high quality factors, low mode volumes, and high coupling efficiencies. However, optimizing across these metrics requires exploring a large design space, motivating the use of inverse design strategies. Previous inverse design efforts targeted high quality factors and low mode volumes, sacrificing the coupling efficiency or lacking the ability to precisely control the far-field radiation pattern. In this work, we present an inverse design framework that simultaneously optimizes cavity quality factor and far-field numerical aperture, both specified as design targets. Using this method, we design L3 photonic crystal cavities with different far-field numerical apertures in the visible-wavelength range and fabricate them in silicon nitride. \rev{Photoluminescence measurements confirm experimental control of the far-field numerical aperture and reveal simultaneous 27.4-fold and 3.4-fold improvements in the coupling efficiency and quality factor, respectively, compared to the standard L3 cavity.} Disorder analysis further shows that the designs retain significant performance despite nanofabrication imperfections. Our work demonstrates a versatile inverse design framework for multi-objective optimization of photonic crystal cavities to attain high quality factors and coupling efficiency.

\end{abstract}

\maketitle
 
\section{Introduction}
Photonic crystal cavities confine light to subwavelength volumes, greatly enhancing light-matter interactions, which has driven advances in low-threshold nanolasers \cite{ellisUltralowthresholdElectricallyPumped2011, yangRoomTemperatureContinuouswave2017}, nonlinear optics \cite{martyPhotonicCrystalOptical2021}, and sensing \cite{zhangReviewOpticalSensors2015}. In the context of cavity quantum electrodynamics, coupling a quantum emitter to a photonic crystal cavity enables a wide range of functionalities including single and entangled photon generation \cite{lodahlInterfacingSinglePhotons2015a, chopinUltraefficientGenerationTimeenergy2023}, spin-photon interfaces \cite{singhOpticalTransparencyInduced2022, luoSpinPhotonInterface2019, sipahigilIntegratedDiamondNanophotonics2016}, single-photon nonlinearity \cite{desantisSolidstateSinglephotonFilter2017, biswas_single-photon_2026}, and single-photon gates \cite{sunQuantumPhaseSwitch2016}. Realizing these applications requires cavities that simultaneously achieve high quality factors, low mode volumes and efficient optical coupling. However, the design space for photonic crystal cavities grows exponentially with the number of parameters, making manual optimization impractical and motivating different design approaches such as inverse design.

\begin{figure*}
    \centering
    \includegraphics[width=\linewidth]{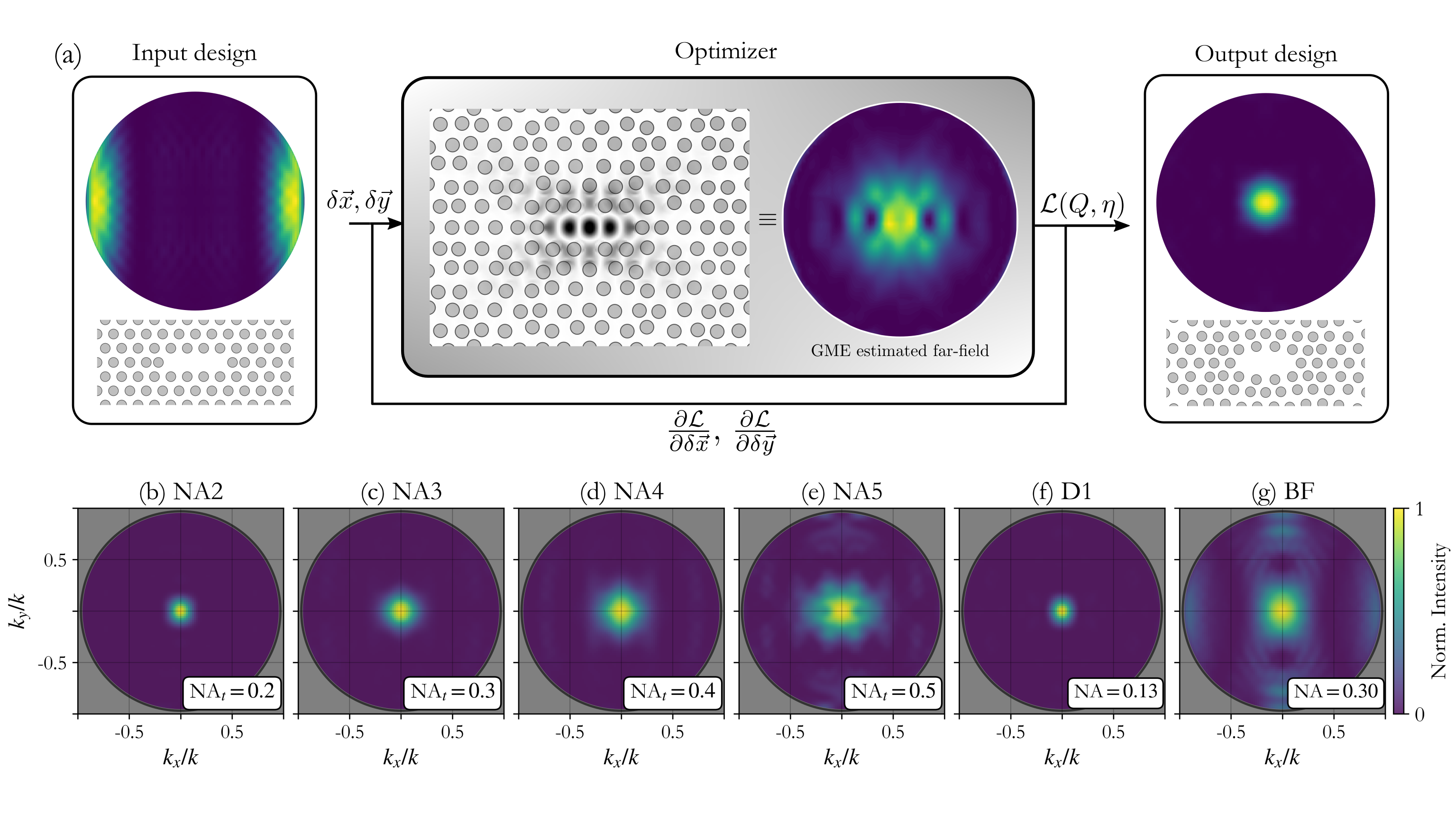}
    \caption{\rev{\textbf{Inverse design of L3 photonic crystal cavity.} (a) Starting with an initial L3 cavity design, the optimizer uses the guided mode expansion method to estimate the cavity quality factor $Q$ and the coupling efficiency $\eta$ of the far-field radiation pattern. These outputs are used to compute the loss function $\mathcal{L}(Q, \eta)$, whose minimization results in the output cavity design with quality factor $Q_t$ and a far-field radiation pattern with numerical aperture $\text{NA}_t$, set during optimization. Hole displacements for the output designs are exaggerated ($10\times$) for visualization. (b)--(e) Far-field radiation patterns, calculated using FDTD, of four inverse-designed cavities, labeled \rev{NA2, NA3, NA4, and NA5}, that exhibit a Gaussian-like far-field with the target numerical aperture indicated in the each panel. Far-field radiation patterns of the directional (f) and the band folding (g) designs with the subtext indicating the best-fit numerical aperture (see main text for details).}}
    \label{fig_invd_framework}
\end{figure*}

Inverse design is a powerful tool for the optimization of photonic crystal cavities because it aims to directly determine the structure that achieves the target performance \cite{maDeepLearningDesign2021}. To efficiently traverse the huge design space, genetic algorithms \cite{minkovAutomatedOptimizationPhotonic2014}, neural networks \cite{asanoOptimizationPhotonicCrystal2018, abeOptimizationH0Photonic2020, liDeepLearningbasedModeling2021} and gradient-based approaches \cite{minkovInverseDesignPhotonic2020, dasOptimizationCircularCavities2026} have been used to optimize the cavity quality factor. \rev{Among these methods, gradient-based optimization is particularly well suited when optimizing over a large number of parameters (see Appendix \ref{app_sec:method_comp} for details)}. Nevertheless, designing cavities that simultaneously optimize the quality factor and tailor the far-field radiation pattern is more challenging. Heuristic methods such as band folding improve directionality at the expense of the quality factor \cite{tranDirectiveEmissionHigh2009, tranVerticalHighEmission2010a, portalupiPlanarPhotonicCrystal2010}. Recently, a directional cavity design was demonstrated \cite{panuskiFullDegreeoffreedomSpatiotemporal2022, saggioCavityenhancedSingleArtificial2024a} that used gradient-based inverse design to co-optimize the quality factor while maximizing radiation into a collimated mode. However, several applications demand finer control of the far-field, such as engineering a Gaussian profile with a specified numerical aperture that can mode-match the cavity emission to fibers or nanophotonic waveguides. Achieving such precise control of the cavity far-field emission remains a significant challenge.

In this work, we develop an inverse design framework to optimize photonic crystal cavities resulting in simultaneous high quality factors and a controllable numerical aperture in the far field. \rev{Our optimization routine, which only adjusts the position of the holes, avoids the need to finely control the hole diameters, which is extremely challenging for sub-wavelength feature sizes. As a result, the final cavity designs remain within the standard nanofabrication capabilities.} We design four distinct L3 photonic crystal cavities, each with a different target NA in the visible wavelength range. The inverse-designed cavities achieve more than an order of magnitude higher simulated quality factor as compared to previous state-of-the-art L3 cavities under similar conditions. We fabricate the cavities in silicon nitride (SiN) and experimentally benchmark the performance by performing a statistical analysis of the cavity photoluminescence spectra against standard L3, band-folded L3, and directional L3 cavity designs. Our results demonstrate the controllability of the far-field NA. \rev{Moreover, the mode-matched cavity achieves a 27.4-fold improvement in coupling efficiency and a 3.4-fold increase in quality factor compared to the initial L3 design, and also outperforms other cavity designs.} Finally, we use the SEM images of the fabricated structures to estimate the hole radius disorder. By performing disorder analysis, we show that our inverse-designed cavities retain significant performance despite nano-scale imperfections. This work presents an important step towards designing photonic crystal cavities that simultaneously exhibit high quality factors and coupling efficiency. 

\section{Inverse design of L3 Photonic crystal cavity\label{sec:inv_des}}

We use the Guided Mode Expansion (GME) method with a gradient-based approach to perform inverse design of photonic crystal cavities. Conventional simulation techniques such as finite-difference time-domain (FDTD) simulations are computationally intensive, making them impractical for iterative optimization over large design spaces. In contrast, GME is a semi-analytical method that utilizes eigenmodes of a homogeneous permittivity slab to approximate the photonic crystal cavity mode (see Appendix \ref{app_sec:GME} for details) \cite{minkovInverseDesignPhotonic2020, andreaniPhotoniccrystalSlabsTriangular2006}. For our GME simulations, we use the open-source python package $\texttt{Legume}$ \cite{minkovInverseDesignPhotonic2020, zanottiLegumeFreeImplementation2024}, which is also compatible with automatic differentiation. This allows for an efficient computation of gradients of an objective function composed of GME-based outputs with respect to the input parameters. Inverse design makes it possible to simultaneously optimize both the cavity quality factor and the far-field numerical aperture, addressing a long-standing limitation of heuristic or single-objective design strategies.

In Figure \ref{fig_invd_framework}a, we illustrate our inverse design framework for optimizing photonic crystal cavities. Starting from an initial cavity design at the target resonant wavelength, we iteratively optimize the positional displacements of holes within the first quadrant along the x and y axes, given by $\delta \vec{x}$ and $\delta \vec{y}$, respectively. The displacements are subsequently applied symmetrically to holes in the other quadrants to enforce symmetry. Since fabrication of holes of different sizes in the same device is challenging, we exclude the hole radius from the optimization parameters. In each iteration during the optimization process, the optimizer uses the GME method to estimate the cavity mode, its quality factor $Q$, and the far-field radiation profile. These quantities are combined into a loss function $\mathcal{L}$, whose gradients with respect to the optimization parameters are then used to update the hole displacements. The optimization process is repeated until convergence, resulting in a non-intuitive cavity design that minimizes the cost function while satisfying nanofabrication constraints.

By choosing an appropriate cost function, our inverse design approach seeks to simultaneously optimize the quality factor $Q$ and the coupling efficiency $\eta$. The parameter $\eta$ is defined as the overlap integral between the transverse far-field mode of the cavity and an outgoing Gaussian beam characterized by its numerical aperture (NA) (see Appendix \ref{app_subsec:coupling efficiency} for details). \rev{It achieves a simultaneous enhancement of the quality factor and coupling efficiency by reducing the total out-of-plane radiation loss while shaping the remaining outgoing radiation into a Gaussian-like emission profile (see Appendix \ref{app_subsec:physical_mech} for details).} We define the following cost function:

\begin{equation}
    \label{eq_loss_function}
    \begin{aligned}
    \rev{\mathcal{L}(Q, \eta)} & \rev{= \Bigg(\frac{\pi}{4} - \arctan\Big(\frac{Q}{Q_{\text{t}}} \Big)\Bigg)^2 + \Big(1 - \eta(\text{NA}_{\text{t}}) \Big)^2} \\
    & \rev{+ \mathcal{W}\big(E_{FF}, E_G \big)}
    \end{aligned}
\end{equation}

\noindent Here, $Q_t$ and $\text{NA}_t$ are the target quality factor and numerical aperture, respectively, \rev{while $E_{FF}$ and $E_G$ represent the cavity far-field and target Gaussian electric field, respectively.} The $\arctan$ function in the first term normalizes the contribution of $Q/Q_{\text{t}}$, ensuring that the two terms remain comparable in value and preventing the optimizer from disproportionately favoring one over the other. \rev{The third term, $\mathcal{W}\big(E_{FF}, E_G \big)$, acts as a spatial penalty mask in reciprocal space and penalizes far-field emission that exceeds the target NA, forcing the optimizer to suppress high-angle radiation and converge toward a single Gaussian-like lobe with high coupling efficiency (see Appendix \ref{app_subsec:spatial penalty mask} for details).} The optimizer works to minimize the loss function which occurs when $Q$ tends to $Q_{\text{t}}$, $\eta$ tends to unity, \rev{and $\mathcal{W}$ tends to zero}. Using this approach enables simultaneous optimization of the cavity quality factor and the far-field radiation pattern by setting appropriate values of $Q_\mathrm{t}$ and $\eta(\text{NA}_{\text{t}})$, respectively.

For our calculations, we focus on visible-wavelength photonic crystal cavities in SiN. Designing nanophotonic devices in low-index materials is challenging due to weak light confinement, which leads to higher optical losses. Compared to longer wavelengths, visible-wavelength devices require small feature sizes that reach the tolerance limits of conventional nanofabrication techniques. We overcome both of these challenges by designing L3 photonic crystal cavities with a resonance wavelength of $\approx 520$ nm in SiN, which has a low refractive index of $\approx 2$. The initial L3 cavity design has a membrane thickness of $225$ nm, lattice constant $a = 208$ nm, hole radius $= 0.3a$, and the closest hole to the cavity is shifted away by $0.26a$ for gentle confinement of the cavity mode \cite{akahaneHighQPhotonicNanocavity2003}. The quality factor of the cavity is $1300$. 

We leverage the ability to design cavities with controllable far-field NA to design \rev{four} L3 photonic crystal cavities with varying NA$_\mathrm{t}$ and plot their far-field profiles in Figure \ref{fig_invd_framework}b--e. The designs are obtained by running four instances of the inverse design optimization and setting different values for NA$_t$ and $Q_t$ in the cost function. In particular, we choose \rev{\{NA$_t$, $Q_t$\} = \{$0.2, 10^5$\}, \{$0.3, 10^5$\}, \{$0.4, 10^5$\}, and \{$0.5, 10^5$\} and label the cavity designs NA2, NA3, NA4, and NA5, respectively (see Appendix \ref{app_sec:inv_des} for details)}. Since GME is an approximate method, we validate the final designs using first-principles FDTD simulations \cite{flexcomputeFastModernPhotonic} and plot the cavity far-field profiles in Figure \ref{fig_invd_framework}b--e. \rev{The far-field profiles are Gaussian-like, and the coupling efficiencies $\eta_0$, calculated using FDTD simulations, are $> 0.6$ for NA2, NA3, and NA4, and 0.37 for NA5. The quality factors $Q_0$ for the designs NA2, NA3, NA4, and NA5 are $5.6 \times 10^4$, $4.4 \times 10^4$, $4.5 \times 10^4$, and $6.1 \times 10^4$, respectively, which differ slightly from the target quality factors. The coupling efficiencies below unity are attributed to the resolution of the GME simulations (see Appendix \ref{app_subsec: limitation of GME} for details).}

\begin{table*}[t]
    \centering
    \renewcommand{\arraystretch}{1.4} 
    
    \begin{tabular*}{\textwidth}{@{\extracolsep{\fill}} l c c c c c c @{}}
        \toprule
        \makecell{Cavity \\ design} & 
        \makecell{Target Quality \\ Factor $Q_t$} &
        \makecell{Target / Best Fit \\ Far-field NA} &
        \makecell{Quality \\ Factor $Q$} & 
        \makecell{Mode Volume \\ $V/(\lambda/n)^3$} & 
        \makecell{Purcell Factor \\ $F = \frac{3}{4 \pi^2} \Big(\frac{\lambda}{n} \Big)^3 \frac{Q}{V}$} & 
        \makecell{Coupling \\ Efficiency $\eta_0$} \\
        \midrule
        L3  & -      & NA = 1.00       & $1.3 \times 10^3$ & 1.02 & 97   & 0.06 \\
        BF  & -      & NA = 0.30       & $5.0 \times 10^2$ & 1.15 & 33   & 0.29 \\
        D1  & $10^5$ & NA = 0.13       & $8.1 \times 10^4$ & 1.29 & 4772 & 0.60  \\
        NA2 & $10^5$ & NA$_{t}$ = 0.20 & $5.6 \times 10^4$ & 1.38 & 3085 & 0.61 \\
        NA3 & $10^5$ & NA$_{t}$ = 0.30 & $4.4 \times 10^4$ & 1.40 & 2388 & 0.67 \\
        NA4 & $10^5$ & NA$_{t}$ = 0.40 & $4.5 \times 10^4$ & 1.40 & 2451 & 0.69 \\
        NA5 & $10^5$ & NA$_{t}$ = 0.50 & $6.1 \times 10^4$ & 1.43 & 3242 & 0.37 \\
        \bottomrule
    \end{tabular*}
    \caption{Performance metrics for different photonic crystal cavity designs as calculated from FDTD simulations. For the inverse-designed cavities NA2–NA5, the listed numerical aperture corresponds to the prescribed target NA$_t$, whereas for the D1, BF, and L3 designs, it corresponds to the best-fit numerical aperture obtained from the overlap with a Gaussian mode.}
    \label{table:cavity_metrics}
\end{table*}

We benchmark the performance of our inverse-designed cavities with the directional cavity design \cite{panuskiFullDegreeoffreedomSpatiotemporal2022, saggioCavityenhancedSingleArtificial2024a}, the band folding design \cite{tranDirectiveEmissionHigh2009}, and the current state-of-the-art L3 cavity designs under similar conditions \cite{barthModificationVisibleSpontaneous2007, murshidyOpticalPropertiesHybrid2010, adawiOpticalNanocavityIncorporating2010}. We use our inverse design approach with a modified cost function to design a directional cavity with $Q_t$ set to $10^5$ and that we denote as D1 (see Appendix \ref{app_subsec: D1 loss function} for details). The far-field profile for the design as calculated from FDTD simulations is shown in Figure \ref{fig_invd_framework}f. The D1 design has a quality factor of $8.1 \times 10^4$ and the far-field radiation pattern exhibits a highly directional emission. The higher quality factor for the cavity design D1 than for NA2--NA5, which have the same $Q_t$, is an artifact of the approximate nature of the GME method. The NA of the D1 design far-field, estimated from the best-fit Gaussian via an overlap integral, is $\approx 0.13$, and the corresponding coupling efficiency $\eta_0 = 0.60$. The band folding design, labeled as BF, also exhibits a directional far-field radiation pattern as shown in Figure \ref{fig_invd_framework}g, with the best-fit NA $\approx 0.30$ and $\eta_0 = 0.29$. However, this comes at the cost of the quality factor, which drops from the initial value of $1300$ to $500$. \rev{We summarize the performance of the different cavity designs, as obtained from FDTD simulations, in Table \ref{table:cavity_metrics}. Compared to the traditional cavity designs, L3 and BF, the inverse-designed cavities exhibit a higher coupling efficiency and a quality factor that is more than an order of magnitude larger. While the mode volume of the inverse-designed cavities increases by $\sim 1.4$-fold, the substantial improvement in the quality factor leads to up to 49-fold higher Purcell factors compared to the L3 cavity, thereby demonstrating a major advancement over traditional cavity architectures.} Finally, our inverse-designed cavities also have more than an order of magnitude higher simulated quality factor than previously reported L3 cavities and more than three times that of a highly engineered heterostructure cavity in SiN in the visible wavelength range \cite{barthEmissionPropertiesHighQ2008}.

\section{Fabrication and Testing of inverse-designed cavities}

To experimentally validate the improved performance of our inverse-designed cavities, we fabricate the cavity designs NA2--NA5, along with the band folding (BF), directional (D1), and initial L3 cavity (L3) designs for comparison on the same sample. First, we deposit a 225 nm layer of SiN by stoichiometric low-pressure chemical vapor deposition (LPCVD) on a Si substrate with a 2 $\mu$m sacrificial $\text{SiO}_2$ layer on top. We then pattern the cavity designs on SiN using standard electron beam lithography, followed by fluorine-based inductively coupled plasma reactive dry etching. Finally, we suspend the patterns by removing the sacrificial $\text{SiO}_2$ layer using chemical wet etching with a buffered oxide solution \cite{purkayasthaPurcellEnhancedEmission2024}.

\begin{figure}[hbt]
    \centering
    \includegraphics[width=\linewidth]{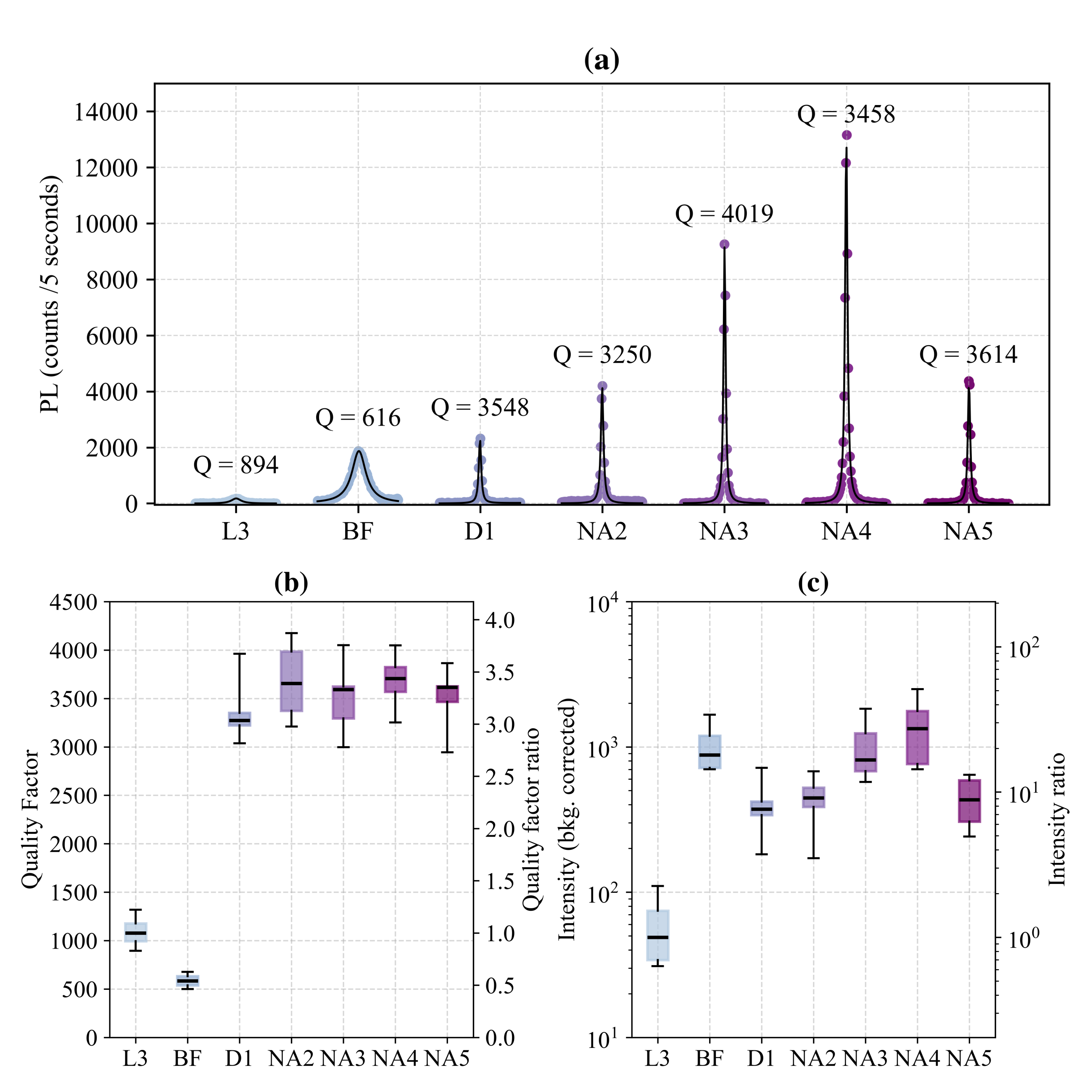}
    \caption{\rev{\textbf{Experimental demonstration of controllable far-field NA.} (a) Photoluminescence spectra for different cavity designs measured at the same power. Because of the Gaussian-like far-field radiation patterns and better mode matching, the inverse-designed cavities NA2--NA5 exhibit higher intensities than initial L3 and directional (D1) cavity designs. The observed trend in the intensity arises due to the different far-field NA of the cavity designs. The mode matching between the NA4 design and the collection optics yields the highest coupling efficiency among all tested designs. Statistical measurements of the quality factor (b) and background-corrected intensity (c) reveal the same trends, verifying that the performance improvements of the inverse-designed cavities are not limited to isolated devices. The NA4 design exhibits a 3.4-fold improvement in the quality factor and a 27.4-fold improvement in coupling efficiency.}}
    \label{fig:experimental_result}
\end{figure}

We characterize the fabricated cavities using the broadband intrinsic photoluminescence of SiN \cite{kistnerPhotoluminescenceSiliconNitride2011}. We perform our measurements in a confocal microscope setup at room temperature. The sample is excited with an above-band laser at $405$ $\mathrm{nm}$, and the cavity emission is collected using an objective lens with an NA of $0.55$. A notch filter at $405$ $\mathrm{nm}$ suppresses the reflected above-band laser light, and the filtered cavity emission is coupled to a single-mode fiber (NA = $0.1$) using a focusing lens, which together form the collection optics. The cavity photoluminescence is recorded using a grating spectrometer with a spectral resolution of 0.02 nm. For the collection optics in our experimental setup, a Gaussian beam with NA = $0.29$ results in optimal mode matching. \rev{The NA at which $\eta$ reaches its maximum does not exactly coincide with the target value $\text{NA}_{t}$. This discrepancy arises because the optimizer maximizes the overlap integral to ensure a large value of $\eta$(NA$_t$), which does not necessarily correspond to the global maximum of $\eta$. As a result, the NA that yields the highest $\eta$ in practice is slightly smaller than $\text{NA}_{t}$. For the ideal, disorder-free cavity designs, we note that NA4 ($\text{NA}_{t} = 0.4$) is better mode-matched with the collection optics than NA3 ($\text{NA}_{t} = 0.3$), followed by NA5, D1, NA2, and BF (see Appendix \ref{app_subsec:coupling efficiency} for details).}

In Figure \ref{fig:experimental_result}a, we show the background-corrected photoluminescence emission of the fundamental mode from different cavity designs, measured under identical excitation power. The background-correction is performed by subtracting the constant and linear background from the Lorentzian cavity response. \rev{As a measure of the coupling efficiency, we use the background-corrected intensity, which is proportional to the number of photons emitted by the cavity that are collected via the collection optics.} Fitting the emission spectra to Lorentzian profiles allows us to extract both the quality factor and the intensity for each design. The measured quality factors are much smaller than the simulated values primarily due to fabrication imperfections (see Section \ref{sec:disorder} for details). Nevertheless, they follow the same relative trend as the simulated quality factors. \rev{The measured quality factor for the cavity design NA5 exceeds the previous state-of-the-art quality factors for L3 photonic crystal cavities \cite{barthModificationVisibleSpontaneous2007, murshidyOpticalPropertiesHybrid2010, adawiOpticalNanocavityIncorporating2010}, and a highly engineered heterostructure cavity \cite{barthEmissionPropertiesHighQ2008} fabricated in SiN.}

\rev{The photoluminescence spectra for our inverse-designed cavities NA2-NA5 exhibit much higher integrated intensity, and thus coupling efficiency, as compared to other cavity designs.} This improvement is due to the Gaussian-like far-field radiation profiles that result in better mode-matching with the collection optics. Importantly, the trend in intensity among the cavity designs, with NA4 performing best, directly reflects the different far-field NA engineered during the optimization process. This establishes clear experimental evidence that the numerical aperture of the cavity far-field can be controllably tailored using inverse design. The huge NA mismatch between the directional cavity design D1 (NA = 0.13) and our collection optics (NA = 0.29) greatly reduces its coupling efficiency, resulting in the lowest coupling efficiency among the optimized designs. \rev{Furthermore, while the band-folding design has improved coupling efficiency compared with the L3 cavity, the inverse-designed cavities exhibit substantially higher quality factors, and the mode-matched NA4 design also achieves a higher coupling efficiency. We note that, contrary to our ideal theoretical calculations, the BF design outperforms the cavity designs NA2, NA3, and NA5. This discrepancy can be accounted for by including the effects of imperfect sidewall angles (see Section \ref{sec:disorder} for details).}

To verify that the high performance of our inverse-designed cavities is not limited to isolated devices, we performed statistical measurements across 8--9 cavities for each design. Figures \ref{fig:experimental_result}b and \ref{fig:experimental_result}c show box plots of the measured quality factors and background-corrected integrated intensity, respectively, for the different cavity designs. In these plots, the boxes extend from the first to the third quartile, the black lines represent the medians, and the whiskers mark the data extremes. Both plots reflect the trends observed in Figure \ref{fig:experimental_result}a, demonstrating that the improvements in quality factor and far-field coupling are robust and reproducible. \rev{The measured median quality factors for NA2--NA5 range from 3200 to 3700, representing $\sim$3.4$\times$ and $\sim$6.3$\times$ improvements over the initial L3 and BF cavity designs, respectively. Furthermore, the NA2, NA3, NA4, and NA5 designs demonstrate coupling efficiency improvements of 9.1$\times$, 16.6$\times$, 27.4$\times$, and 8.8$\times$, respectively, compared to the L3 cavity. By contrast, the BF and D1 designs show improvements of 18$\times$ and 7.6$\times$, respectively. Notably, the BF design has an extremely poor quality factor and, consequently, a large linewidth. This broad linewidth yields integrated intensity values comparable to NA4, even though NA4 possesses a narrower linewidth and significantly higher peak counts.} Together, these results provide experimental validation that our inverse design approach yields cavities with a controllable far-field NA that simultaneously achieve high quality factors.

\begin{figure}
    \centering
    \includegraphics[width=\linewidth]{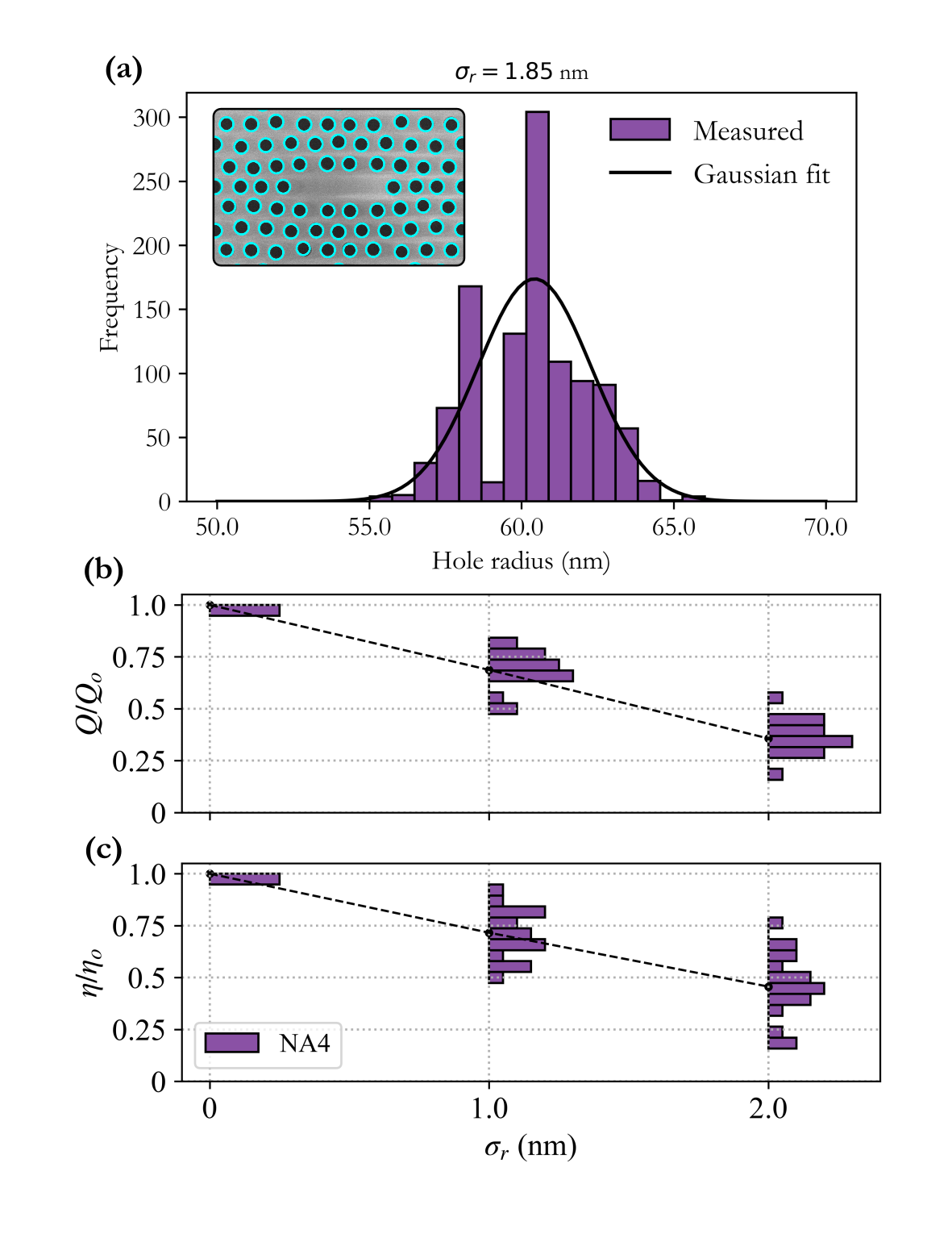}
    \caption{\rev{\textbf{Disorder analysis of inverse-designed cavity.} (a) Histogram of hole radii obtained by an image-processing algorithm that identifies hole contours (inset) from the SEM image of the photonic crystal cavity. Fitting the histogram data with a Gaussian distribution reveals a standard deviation $\sigma_r = 1.85$ nm. By running several instances of disordered photonic crystals in FDTD for different disorder strengths $\sigma_r$, we accumulate and plot the distributions of the ratio of the quality factor (b) and coupling efficiency (c), normalized to their respective design values. The black dots indicate median values, and the dashed lines act as a guide to the eye.}}
    \label{fig:noise_analysis}
\end{figure}

\section{\label{sec:disorder} Disorder analysis of inverse-designed cavities}

A major limitation in the performance of photonic crystals is random disorder arising from the nanofabrication process. Since photonic crystals work on the principle of periodically varying the refractive index, disorder in the hole position and radius will reduce the quality factor and the coupling efficiency. It is therefore essential to evaluate how disorder degrades the performance of our inverse-designed cavities, particularly to assess whether it accounts for the discrepancy between simulated and measured quality factors. The effect of disorder in photonic crystal cavities has also been extensively studied both theoretically \cite{minkovEffectHoleshapeIrregularities2012, minkovStatisticsDisorderinducedLosses2013, geraceEffectsDisorderPropagation2005, haginoEffectsFluctuationAir2009} and experimentally \cite{portalupiDeliberateIntrinsicDisorder2011, taguchiStatisticalStudiesPhotonic2011}. Because our electron-beam lithography offers high positional accuracy, we neglect hole position disorder and consider radius disorder as the dominant error source. 

We estimate the disorder in our nanofabrication process using an image-processing algorithm that analyzes the SEM images of the fabricated photonic crystal cavities. The algorithm uses the \texttt{OpenCV} toolbox to accurately identify the hole contours and fits them to circles (see inset of Figure \ref{fig:noise_analysis}a). For one of the inverse-designed cavities, we plot a histogram of the measured hole radii in Figure \ref{fig:noise_analysis}a and fit it with a Gaussian distribution to extract the standard deviation $\sigma_r = 1.85$ nm of the hole radii. We repeat the image processing on other cavity designs and calculate the average standard deviation to be $\sigma_r \approx 1.8$ nm, which provides an upper bound for the hole radius disorder.

To determine the impact of disorder on the performance of our inverse-designed cavities, we calculate the cavity quality factor and coupling efficiency under simulated disorder. For every hole, the radial disorder is sampled independently from a normal distribution with a standard deviation $\sigma_r$. We generate $30$ instances of disordered photonic crystal cavities for a particular $\sigma_r$ and simulate them using FDTD to extract the quality factor and coupling efficiency for each instance. By calculating the median quality factor and coupling efficiency of the data set, we estimate the performance of a typical disordered photonic crystal cavity. Repeating the above process with increasing $\sigma_r$, we estimate the degradation in the performance of the inverse-designed cavities.

In Figures \ref{fig:noise_analysis}b and \ref{fig:noise_analysis}c, we plot the distributions of the ratios of quality factor and coupling efficiency to their respective design values for cavity design NA4 under increasing disorder strength $\sigma_r$. For $\sigma_r = 0$, both values equal unity. As $\sigma_r$ increases, the median values for both the quality factor and coupling efficiency decrease and their distributions broaden. \rev{For $\sigma_r = 2$ nm, the median values of the quality factor and coupling efficiency drop to approximately 0.35 and 0.45 times their original values, respectively. The other cavity designs show behavior similar to NA4 under disorder, with median values for the quality factor and coupling efficiency falling between 0.3 and 0.6 times their design values. Even with hole-radius disorder accounted for, the experimentally measured quality factor remains several-fold lower than the simulated values, a discrepancy that we attribute to non-vertical sidewall angles. Including the experimentally measured sidewall angle together with hole-radius disorder yields a median simulated quality factor of $4.3\times10^3$, close to the measured value of $3.7\times10^3$ (see Appendix \ref{app_sec:sidewall angle} for details).} Overall, these results demonstrate that our inverse-designed cavities retain significant performance despite nanofabrication disorder.

\section{Conclusion and outlook}

In conclusion,  we present an inverse design approach for multi-objective optimization of photonic crystal cavities to simultaneously achieve high quality factors and controllable far-field numerical aperture. Using this approach, we designed four L3 cavities in the visible wavelength range. \rev{By fabricating these designs in SiN and performing statistical measurements of the cavity photoluminescence spectra, we show  experimental control of the far-field NA, along with a simultaneous $27.4$-fold enhancement in coupling efficiency and a $3.4$-fold increase in quality factor relative to the initial L3 design.} Furthermore, disorder analysis demonstrates that the inverse-designed cavities retain substantial performance even in the presence of nanofabrication imperfections. 

The coupling efficiency of the final designs can be further improved by refining the optimization loss function and placing stricter constraints on the far-field radiation pattern. \rev{Beyond the L3 cavity geometry, our framework can also be used for other hole-based photonic crystal architectures such as H1, L4/3, and heterostructure cavities, making it a versatile optimization tool.} The significant enhancement in coupling efficiency has implications for cavity quantum electrodynamics: quantum emitters can be accessed and controlled with significantly lower excitation powers, reducing laser-induced dephasing and improving spin coherence \cite{bodeyOpticalSpinLocking2019, farfurnikAllOpticalNoiseSpectroscopy2023a}. Efficient photon extraction also accelerates $N$-photon correlation measurements, with the 27.4-fold increase in coupling efficiency translating into a $27.4^{N}$-fold reduction in acquisition time. Importantly, our designs rely  on etching holes of the same radius in a photonic crystal slab, simplifying nanofabrication and improving yield compared to more complex devices that aim to increase the coupling efficiency, such as nanobeam cavities \cite{islamCavityEnhancedEmissionSilicon2024} or nanowire geometries \cite{gregersenControllingEmissionProfile2008, perezDirectLaserWrittenPolymerNanowire2025, munschDielectricGaAsAntenna2013}. Ultimately, this work paves the way for designing optimized photonic crystal cavities that find applications in a broad range of fields such as cavity quantum electrodynamics, optoelectronics, and display technologies. \\

\textbf{Data Availability:} All requests for code and data should be made to Neelesh Kumar Vij at \url{nkvij@umd.edu}. \\

\textbf{Code Availability:} Source code for the simulations and accompanying tutorials can be found on the \href{https://github.com/nkvij/Inverse-design-of-L3-photonic-crystal-cavity}{GitHub repository}. \\

\textbf{Author Contributions:} The inverse design algorithm was developed by N.K.V., J.R.B., and E.W. Detailed simulations and numerical analyses were performed by N.K.V. P.P. fabricated the sample and N.K.V. performed the experiments. All authors contributed to the editing and writing of the manuscript. \\

\textbf{Competing Interest:} The authors declare no competing interests. 

\section*{Acknowledgments}
The simulations presented in this paper were performed on the Zaratan supercomputing cluster. The authors acknowledge the University of Maryland supercomputing resources (http://hpcc.umd.edu) made available for conducting the research reported in this paper. The authors would like to thank Momchil Minkov and Christopher Panuski for helpful discussions.

\section*{Funding}
The authors would like to acknowledge funding support from the National Science Foundation (grants \# UWSC12985 and \# ECCS2423788), the Air Force Office of Scientific Research (grants \# FA95502410266 and \# FA95502210339), and IMOD (\# DMR-2019444). 

\newcounter{keepfigure}
\setcounter{keepfigure}{\value{figure}} 
\appendix 
\setcounter{figure}{\value{keepfigure}} 
\renewcommand{\thefigure}{\arabic{figure}} 

\appendix
\renewcommand{\theequation}{\thesection.\arabic{equation}}
\renewcommand{\thesubsection}{\thesection \arabic{subsection}}

\rev{\section{\label{app_sec:method_comp} Comparison between different optimization algorithms}}

\rev{The computational complexity of photonic crystal optimization depends primarily on the number of cavity mode calculations, which is the most computationally expensive step in the process. Therefore, an efficient optimization strategy must minimize these calculations.}

\rev{Gradient-based approaches coupled with automatic differentiation require only two cavity mode computations per optimization step (a forward and a backward pass), completely independent of the number of optimization parameters. This makes it a highly efficient method for navigating a large design space, as the computational complexity per step does not scale with the dimensionality of the problem. In our approach, the total number of cavity mode computations is on the order of $\sim 10^2$ (twice the number of training epochs).}
    
\rev{In contrast, for alternative inverse design strategies such as neural network-based methods and genetic algorithms, the computational cost scales steeply with the number of optimization parameters. Neural networks require a large upfront training data set to accurately model the underlying physics; previous works have required datasets of $\sim 10^3$--$10^4$ cavity simulations just for training \cite{liDeepLearningbasedModeling2021, asanoOptimizationPhotonicCrystal2018}. Similarly, for genetic algorithms, the number of cavity computations equals the number of generations multiplied by the population size. Past work \cite{minkovAutomatedOptimizationPhotonic2014} required roughly 300 generations of 120 individuals, totaling $>3 \times 10^4$ cavity computations merely to optimize 7 parameters. Because the required number of generations increases rapidly with the number of optimization parameters, these alternative methods quickly become computationally impractical for high-dimensional design spaces. \\}

\section{\label{app_sec:GME} GME method for simulation of
photonic crystal cavity}

Under the assumption of a linear, isotropic, nonmagnetic, and nondispersive medium along with the absence of free charges and currents, Maxwell's equations can be written as an eigenvalue problem for the magnetic field $(\mathbf{H})$ alone:

\begin{equation}
    \label{GME-eigen}
    \Bigg[\nabla \times \frac{1}{\epsilon(\mathbf{r})} \nabla \times \Bigg] \textbf{H(r)} = \frac{\omega^2}{c^2} \textbf{H(r)}
\end{equation}

\noindent where $\epsilon(\textbf{r)}$ is the relative dielectric permittivity of the slab and $\omega$ is the eigen frequency for the system. To solve this eigenvalue problem, the basis chosen by GME consists of the guided modes of an "effective slab" that has homogeneous permittivity. This effective slab has the same thickness as the photonic crystal slab, with its permittivity set to the average of the latter's unit cell. The guided modes of the effective slab do not form a complete basis set, which makes the GME method inherently approximate. However, the GME method has been shown to agree well with FDTD results, allowing for an efficient and accurate computation of the cavity eigenmode \cite{minkovInverseDesignPhotonic2020, andreaniPhotoniccrystalSlabsTriangular2006}. 

The quality factor and the far-field can also be accurately computed within the GME formulation by perturbative coupling of the cavity eigenmode with the continuum of radiative modes that lie above the light line. The perturbative coupling coefficients $c^{(s,p)}_{mn}$ represent the radiative loss rates of the cavity mode along the reciprocal lattice vector $\vec{g}_{mn}/k = (\vec{k} + \vec{G}_{mn})/k = \Bigg[\vec{k} +  2\pi \Big(\frac{m}{L_x}, \frac{n}{L_y} \Big) \Bigg]/k$ for polarizations $s$ $(\equiv$ TE) and $p$ $(\equiv$ TM). Here, $L_x, Ly$ is the length of the photonic crystal slab along the $X$ and $Y$ axis and $k = 2 \pi/\lambda$. Including additional reciprocal vectors $\vec{k}$ that lie within the Brillouin zone increases the resolution of the far-field. The total radiative intensity along a reciprocal lattice vector is given by $c_{mn} = |c^{s}_{mn}|^2 + |c^{p}_{mn}|^2$.\\

\section{\label{app_sec:loss_function} Loss function for gradient-based optimization}

\subsection{Coupling Efficiency Calculation\label{app_subsec:coupling efficiency}}

In order to achieve high coupling efficiency, we maximize the overlap integral between the cavity far-field $\textbf{E}_{\text{FF}}$ and a Gaussian beam in the paraxial limit from the fiber of width $w$ propagating along the $\hat{\textbf{z}}$ direction and polarized along $\hat{\textbf{y}}$. The Gaussian beam is given by $\textbf{E}_{\text{G}} = \sqrt{\frac{2c\mu_o}{\pi w^2}} \exp\Big(- \frac{x^2 + y^2}{w^2} \Big) \hat{\textbf{y}}$ and the associated magnetic field $\textbf{H}_G = \dfrac{1}{c \mu_o} \hat{\textbf{z}} \times \textbf{E}_G = -\sqrt{\frac{2}{\pi c\mu_o w^2}} \exp \Big(- \frac{x^2 + y^2}{w^2} \Big) \hat{\textbf{x}}$. The cavity far-field and the Gaussian beam are normalized such that the power carried by them, $\displaystyle \iint \textbf{Re}[(\textbf{E} \times \textbf{H}^*). d\textbf{S}] = 1$. The coupling efficiency $\eta$ is defined as \cite{munschDielectricGaAsAntenna2013}:

\begin{equation}
    \eta = \left\vert \iint \Big(\textbf{E}_{\text{FF}} \times \textbf{H}_{\text{G}}^*\Big). d\textbf{S} \right\vert^2
    \label{eq:eta}
\end{equation}

\noindent The width $w_o$ of the Gaussian beam that results in mode-matching can be converted to numerical aperture using the formula:

\begin{equation}
\text{NA} = \sin \theta = \sin \Bigg(\arctan \Bigg( \dfrac{w_o}{f_o} \Bigg) \Bigg)    
\end{equation}

\begin{figure}[hbt]
    \centering
    \includegraphics[width=\linewidth]{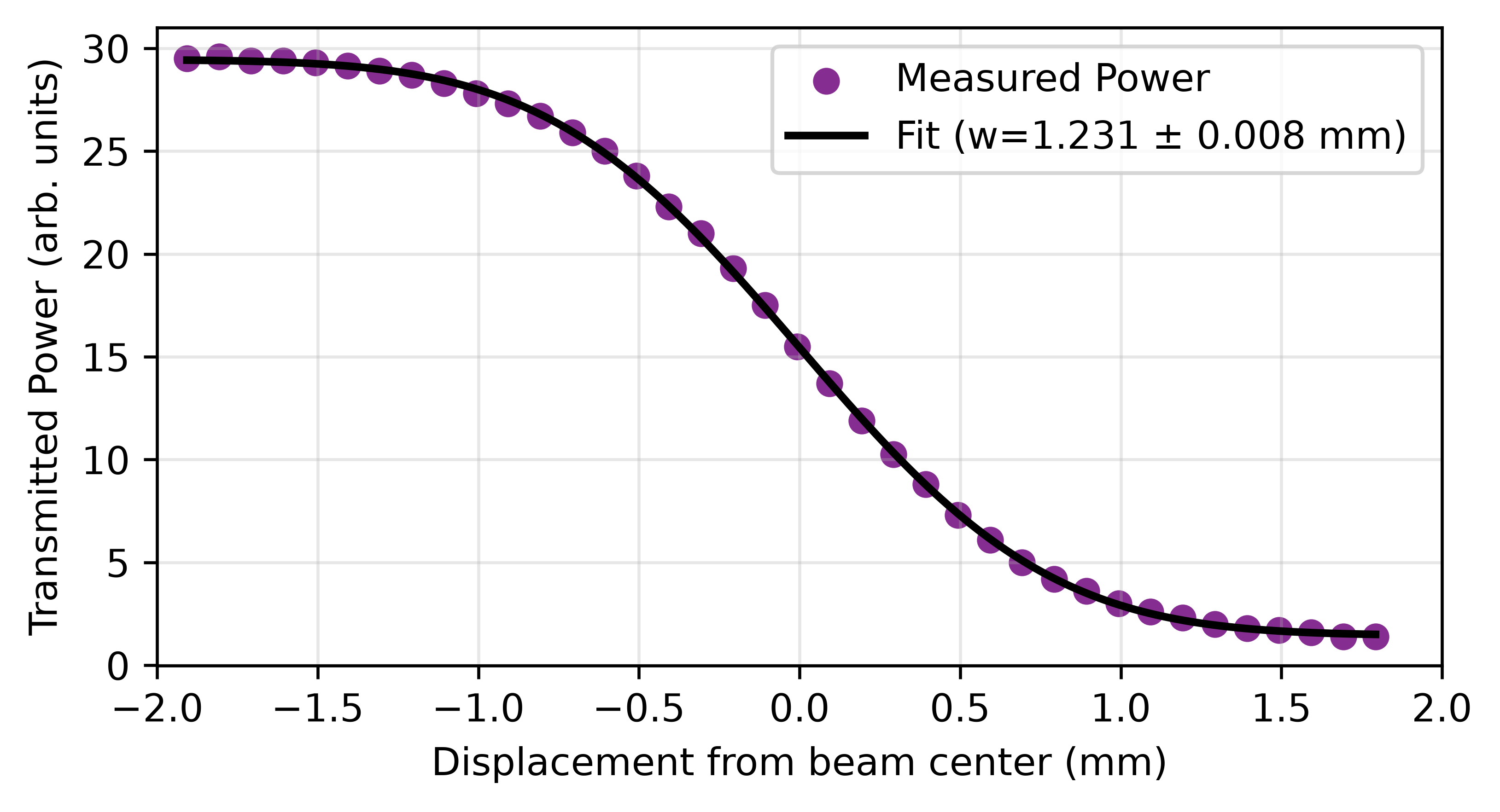}
    \caption{\rev{\textbf{Knife-edge test:} Measured beam power on a power meter as a function of knife displacement. By fitting an error function, we extract a beam width of 1.231 mm for the collection optics, which corresponds to an NA of 0.29.}}
    \label{fig:knife_edge_test}
\end{figure}

\noindent Here $f_\text{o}$ is the focal length of the objective lens. \rev{To estimate the width $w_o$, we perform a knife-edge test. We shine a $520$ nm laser through the collection fiber, collimate the beam with the collection lens, and measure the beam power as a function of knife displacement. In Figure \ref{fig:knife_edge_test}, we plot the measured power as a function of knife displacement and fit an error function to extract the beam's center and width. A beam width of 1.231 mm yields the best fit, which corresponds to an NA of 0.29.}

\begin{figure*}[hbt]
    \includegraphics[width=\linewidth]{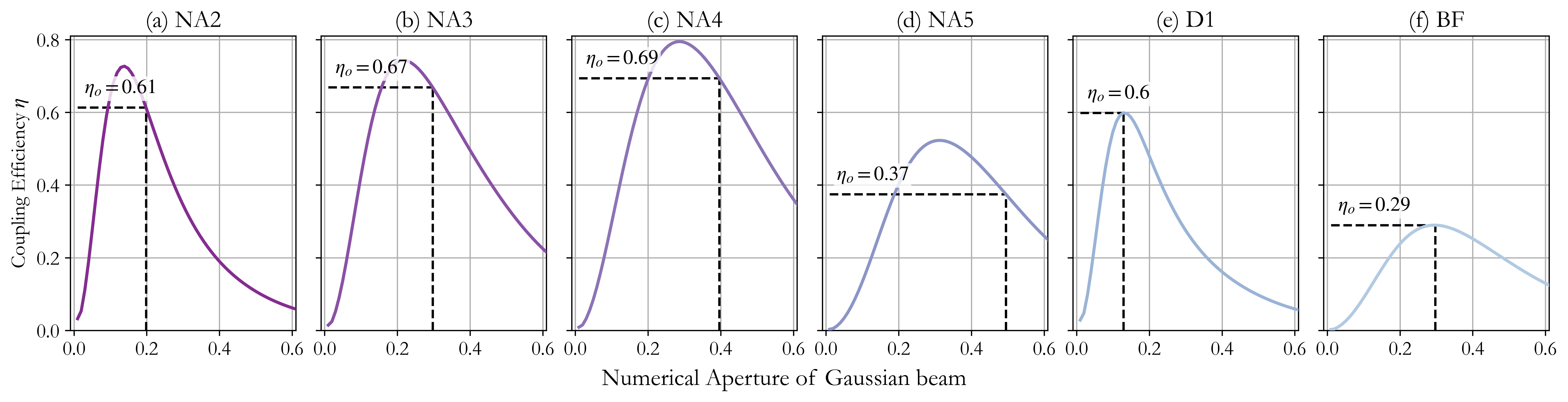}
    \caption{\rev{\textbf{Coupling efficiency for different cavity designs.} We numerically calculate the coupling efficiency using Eq. \ref{eq:eta} as a function of the numerical aperture of the Gaussian beam $\textbf{E}_{\text{G}}$. The calculations are done using the cavity far-fields obtained from FDTD simulations.}}
    \label{fig:gaussian_overlaps}
\end{figure*}

Using Eq. \ref{eq:eta}, we evaluate $\eta$ for different L3 cavity designs, and plot it as a function of the NA of the Gaussian beam $\textbf{E}_{\text{G}}$ in Figure \ref{fig:gaussian_overlaps}. The far-field radiation pattern is calculated by applying a near-to-far-field transformation to the FDTD-simulated cavity mode on a plane positioned above and parallel to the photonic crystal slab \cite{flexcomputeFastModernPhotonic}. \\

\rev{\subsection{\label{app_subsec:spatial penalty mask}Spatial penalty mask}}

\rev{The exact form of $\mathcal{W}\big(E_{FF}, E_G \big)$ is given by:}

\begin{equation}
    \rev{\mathcal{{W}}\big(E_{FF}, E_G \big) = \frac{ \displaystyle \iint \Bigg(1 - \frac{|E_G|}{\max{(|E_G|)}} \Bigg)^2 |E_{FF}|^2 \sin(\theta) d\theta d\phi}{\iint |E_{FF}|^2 \sin(\theta) d\theta d\phi}}
\end{equation}

\rev{The weighting factor $(1 - |E_G|/\max{(|E_G|)})$ acts as an inverted mask of the target Gaussian field, that is zero near the origin and approaches unity at the tails. Squaring this mask provides smoother convergence while maintaining a sufficient boundary penalty to suppress high-angle emission. To compute the penalty for high-NA emission, this mask is multiplied by the cavity far-field intensity $|E_{FF}|^2$, integrated over the solid angle, and normalized by the total far-field intensity. Consequently, the value of $\mathcal{W}\big(E_{FF}, E_G \big)$ is bounded between $[0, 1)$, and the optimizer actively works to minimize it.} \\

\section{\label{app_sec:inv_des} Inverse design based cavity optimization}

\subsection{\label{app_subsec: GME sim setup}GME Simulation setup}

\begin{figure}
    \includegraphics[width=\linewidth]{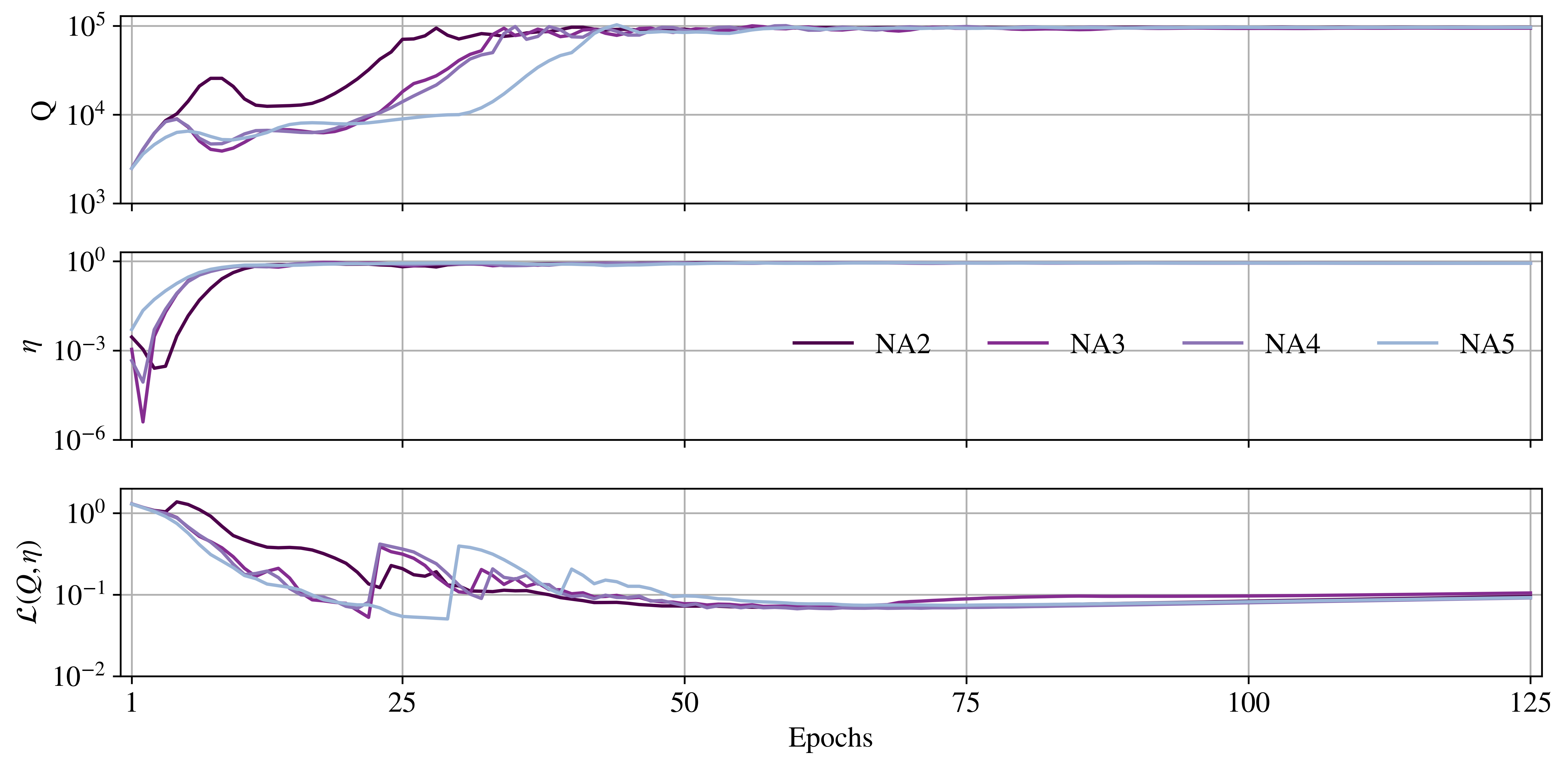}
    \caption{\rev{\textbf{Inverse-design-based cavity optimization.} For the four inverse-designed cavities---NA2, NA3, NA4, and NA5, the GME-calculated quality factor, coupling efficiency, and the loss function are shown as a function of training epochs. All values converge to the target values over the course of $125$ iterations.}}
    \label{fig:training_curves}
\end{figure}

To simulate the cavity quality factor and far-field, we use only the lowest-order guided mode and aggregated the loss rates $c_{mn}$ over a $2 \times 2$ grid of Bloch boundary conditions $\vec{k}$ at all wave vectors $\vec{g}_{mn}$ such that $|\vec{g}_{mn}| < g_{max} = 2.5 \times 2\pi/a$. \rev{For an accurate calculation of the complex electric far-field in GME, a phase correction is necessary for the additional far-field components corresponding to reciprocal lattice vectors $\vec{k}$, as these components inherently possess an arbitrary gauge. We determine this correction by first computing the $y$-component of the cavity far-field and observing that nearby reciprocal lattice vectors, when sampled finely, share the same phase. This relationship provides the necessary phase correction for all additional $\vec{k}$ vectors.}

\subsection{\label{app_subsec: D1 loss function}Loss function for Directional cavity design}

The loss function for the cavity design D1 has only two components, unlike the loss function defined in the Eq \ref{eq_loss_function} of the main text. The quality factor contribution is the same and we define the far-field contribution as $(1 - \eta_d)^2$, where $\displaystyle \eta_d = \dfrac{c_{00}}{\Gamma}; \Gamma = \displaystyle \sum_{mn} c_{mn}$ \cite{panuskiFullDegreeoffreedomSpatiotemporal2022, saggioCavityenhancedSingleArtificial2024a}.

\subsection{\label{app_subsec:adam optimization} Gradient-based optimization simulation setup}

\rev{We choose the L3 cavity as the starting point of our optimization because it inherently has a low mode volume and high baseline quality factor. Consequently, our inverse design algorithm can focus on optimizing small hole displacements to minimize the loss function, rather than having to reinvent an efficient mode confinement mechanism from scratch.}

For our inverse design process, we only consider the position of the $16 \times 16$ array of holes closest to the cavity out of the $32 \times 32$ hole array as optimization parameters. The hole displacements are bounded between ($-0.1 \mathrm{a}, 0.1 \mathrm{a}$) to avoid holes getting too close to each other.  We use the Adam optimizer and perform the training in three stages. In successive stages, we increase the target quality factor from $Q_{\text{target}}/100 \to Q_{\text{target}}/10 \to Q_{\text{target}}$ when the quality factor reaches $0.9$ times the target quality factor in that stage. Furthermore, we exponentially anneal the learning rate to one-tenth its value over the course of optimization. To efficiently keep track of the cavity mode during optimization, we first save the eigenvector (obtained by solving Eq. \ref{GME-eigen}) corresponding to the initial L3 cavity design. During training, we identify the cavity mode by calculating the dot product between the eigenvectors of the intermediate design and the previously saved eigenvector. The eigenvector that results in the highest dot product (typically $\sim 0.9$) represents the correct cavity mode. The course of optimization for the four cavity designs are shown in Figure \ref{fig:training_curves}. Over $125$ epochs, the quality factor and coupling efficiency reach their target values and the loss function drops more than two orders of magnitude, converging to $< 10^{-1}$. The discrete jumps in the plots appear when the optimizer switches from one stage to another.

\rev{\subsection{\label{app_subsec: final cavity designs}Final cavity designs}}

\rev{The different inverse-designed cavities share the same hole diameter (124.8 nm), lattice constant (208 nm), and slab thickness (225 nm). The only difference between the designs is the positions of the holes within the photonic crystal lattice. To better visualize the hole displacements for the inverse-designed cavities, we plot the hole positions for cavity designs NA2--NA5 in Figures 7a--d, respectively. The initial L3 cavity hole positions are represented by black dashed lines, and the final hole positions are colored according to the magnitude of their displacement---the greater the displacement, the darker the color.} \\

\begin{figure}[hbt]
    \centering
    \includegraphics[width=\linewidth]{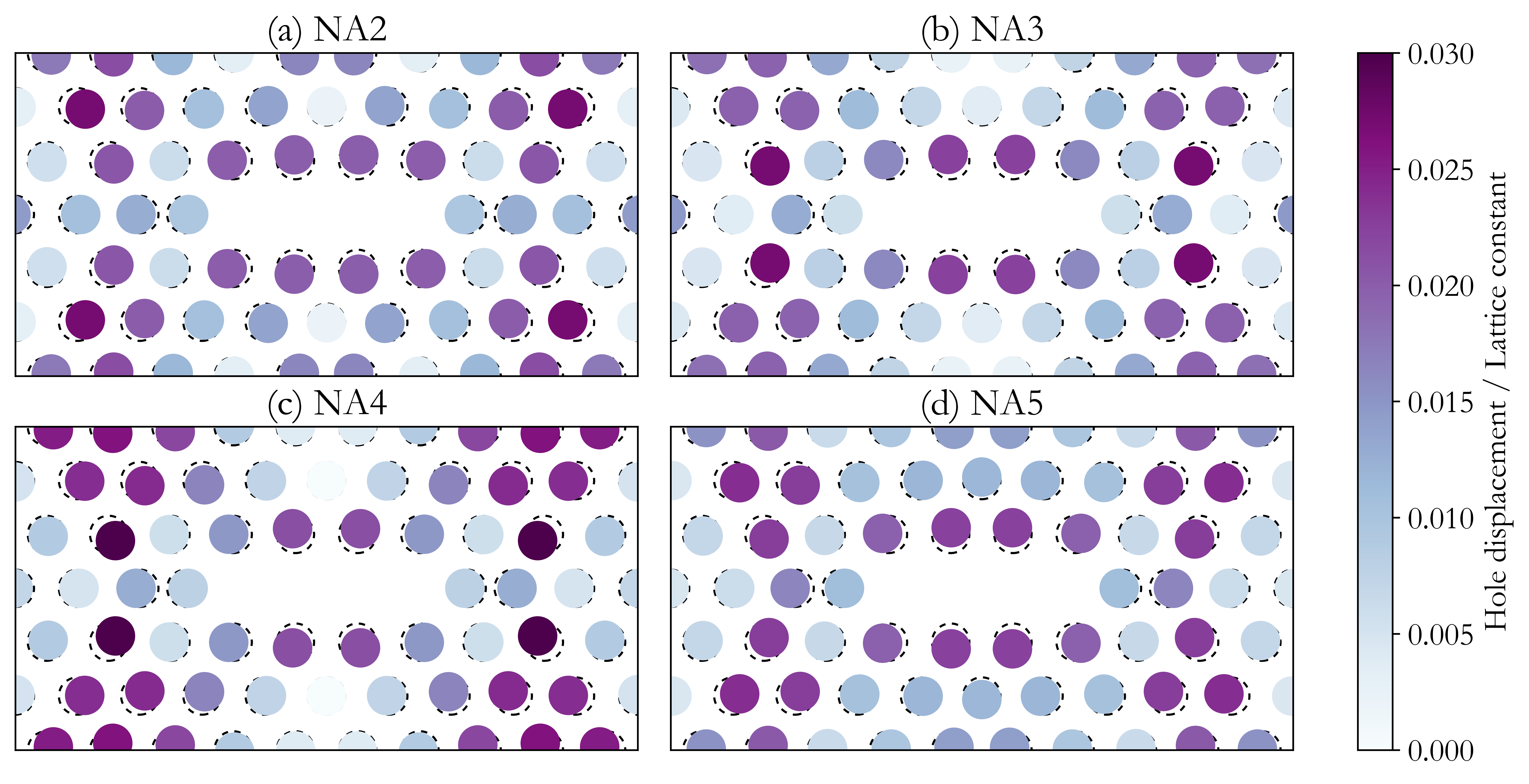}
    \caption{\rev{\textbf{Hole displacements for inverse-designed cavities.} In panels (a)--(d), we plot the hole positions for the cavity designs NA2--NA5. The initial hole positions are indicated by the black dashed lines, and the final holes are colored in proportion to their displacements---the greater the displacement, the darker the color.}}
    \label{fig:hole_displacements}
\end{figure}

\rev{\subsection{\label{app_subsec:intial condition}Effect of initial condition}}

\rev{To investigate the effect of varying initial conditions on the optimization, we reran the inverse design optimizer starting from a perturbed L3 cavity. Instead of starting from the nominal L3 cavity design, we added noise to the initial values of the optimization parameters (hole $x$- and $y$-displacements in the first quadrant). The noise was sampled from a continuous uniform distribution ranging from $[-\sigma_p, \sigma_p]$, where we define $\sigma_p$ as the initial parameter noise. We set $\{\text{NA}_t, Q_t\} = \{0.3, 10^5\}$ and ran our inverse design optimization on 20 randomized initial structures for each value of $\sigma_p$: 5, 10, 20, and 40 nm.}

\rev{To quantify the similarity of the final designs, we computed the normalized dot product between the final parameters of each noisy run and the nominal design (final design for $\sigma_p = 0$). This metric, bounded between [-1, 1], quantifies the overlap between two cavity designs. We selected this metric over traditional Euclidean distance because the latter can be misleading. For example, two designs might differ by small displacements of two different, physically distant holes. While the Euclidean distance between these final parameter sets might be small, the resulting photonic crystal cavities would have very different performances.}

\begin{figure}[hbt]
    \centering
    \includegraphics[width=\linewidth]{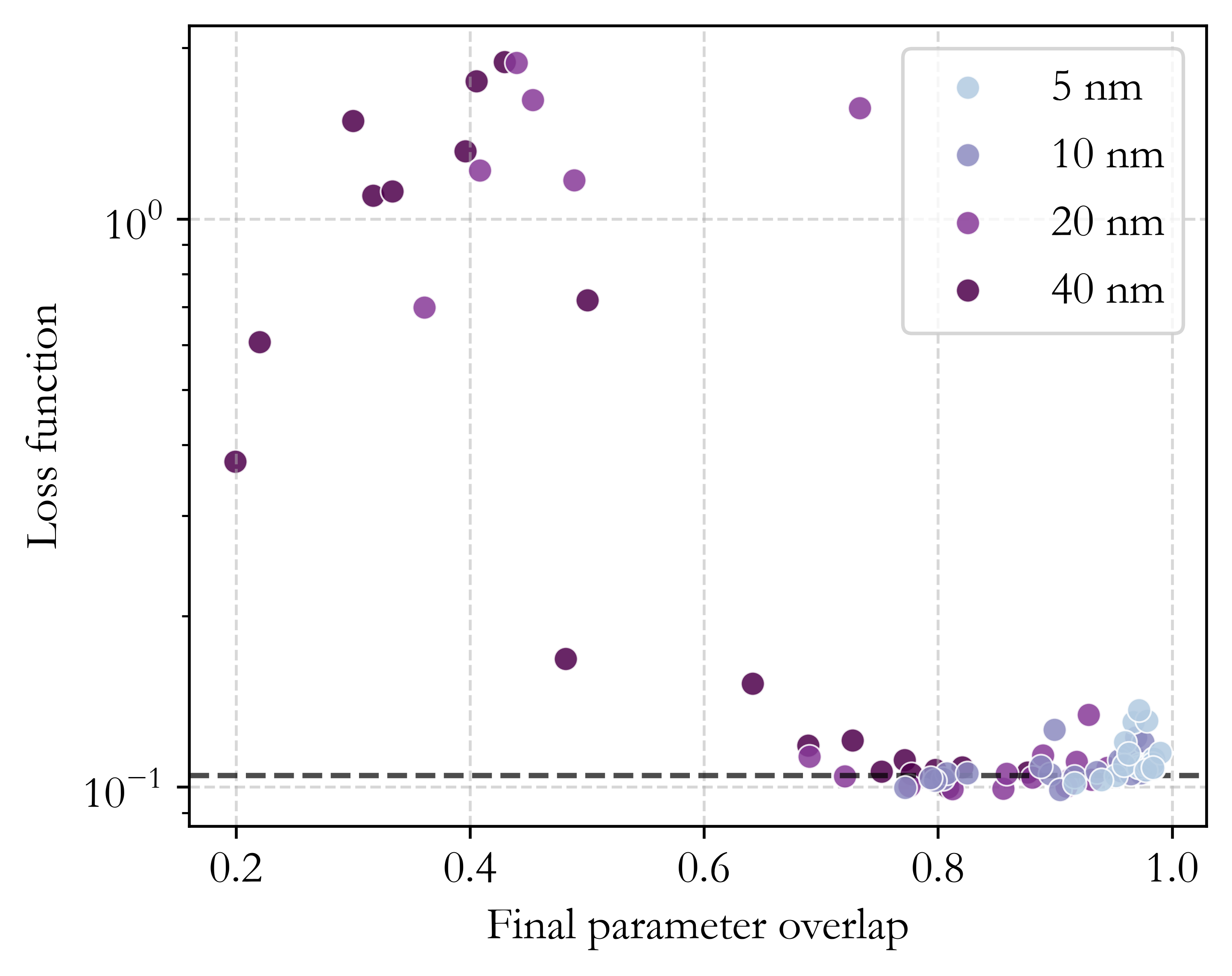}
    \caption{\rev{\textbf{Dependence of optimizer convergence on initial structure.} Scatter plot of the GME-calculated loss function versus final parameter overlap for 80 runs (20 per $\sigma_p$ value).}}
    \label{fig:training_noise}
\end{figure}

\rev{In Figure \ref{fig:training_noise}, we show a scatter plot of the loss function versus final parameter overlap for all 80 runs (20 per $\sigma_p$ value). We observe that all runs for $\sigma_p = 5$ and 10 nm cluster tightly in the bottom right, where the loss function reaches the nominal design value (indicated by the black dashed line) and the final parameter overlap is $\gtrsim 0.8$. Conversely, for $\sigma_p = 20$ and 40 nm, a majority of runs fail to converge---more so for $\sigma_p = 40$ nm---yielding high loss values and small final parameter overlaps (top-left cluster). These results indicate that the L3 cavity and small perturbations around it ($\sigma_p = 5, 10$ nm) are good initial conditions because they result in designs that successfully minimize the defined loss function to $< 10^{-1}$, achieving the target quality factor while approaching near-unity coupling efficiency for the target NA. This clearly demonstrates that optimization failure reliably correlates with a small final parameter overlap with the nominal design, indicating that the optimization outcome is heavily dependent on the initial structure. Furthermore, the optimization reliably reproduces the nominal design for small perturbations to the L3 cavity and becomes increasingly prone to convergence failure as the value of $\sigma_p$ increases.}

\rev{\subsection{\label{app_subsec: limitation of GME} Limitation of the GME method}}

\rev{We investigate the limitations of our inverse design algorithm by running the inverse design optimization for NA$_t$ values ranging from 0.2 to 0.9 in increments of 0.1, while maintaining $Q_t = 10^5$. During the optimization, we use a GME resolution setting of $g_{\text{max}} = 2.5$ and aggregate the losses over a $2 \times 2$ array of reciprocal lattice vectors $\vec{k}$. The optimizer successfully converges for all values of NA$_t$.}

\begin{figure*}
    \centering
    \includegraphics[width=\linewidth]{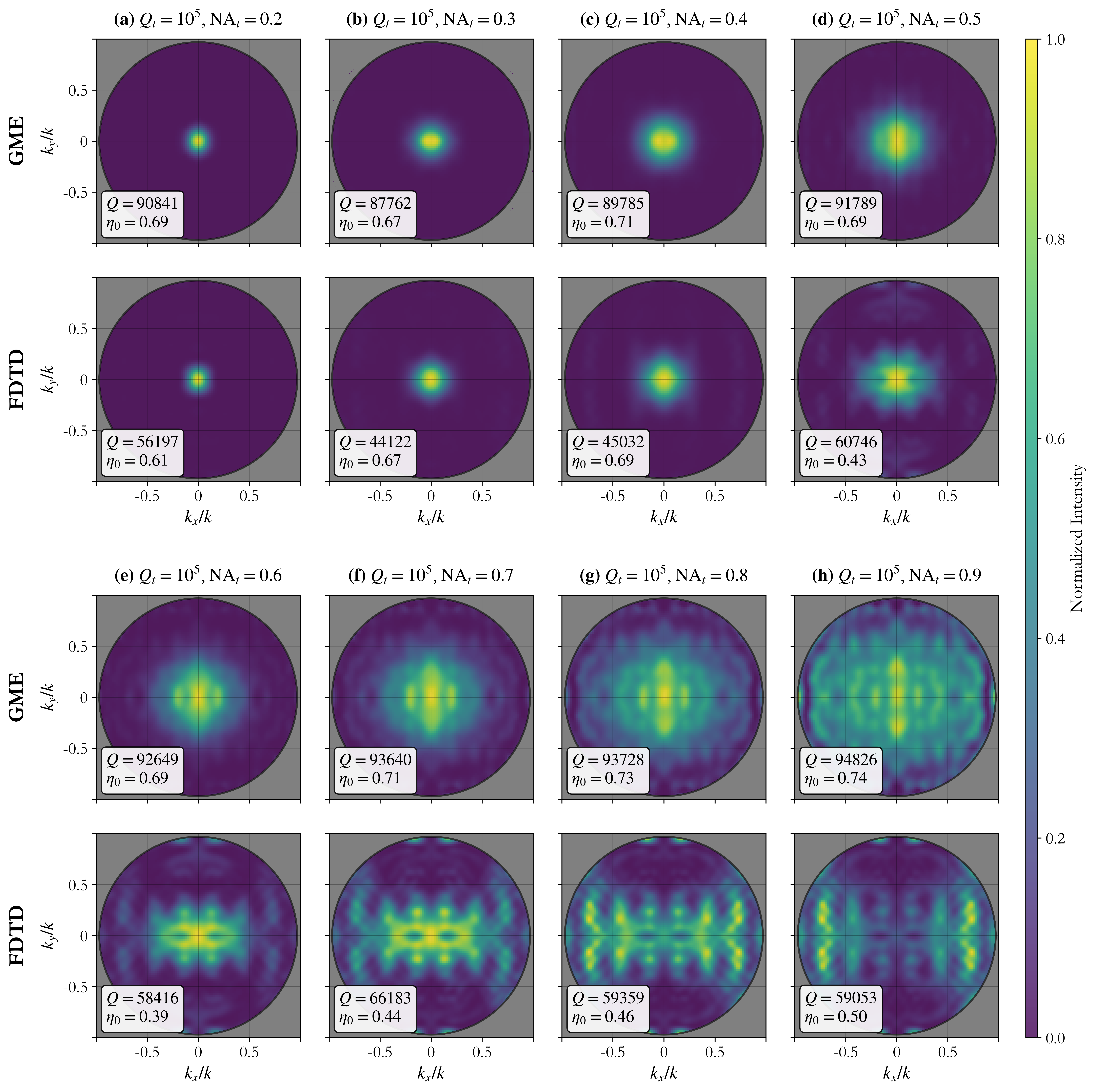}
    \caption{\rev{Performance comparison between the GME (top rows) and FDTD (bottom rows) results for $Q_t = 10^5$ and increasing NA$_{t}$.}}
    \label{fig:GME_FDTD_comparison}
\end{figure*}

\rev{To visualize the final designs obtained from the inverse design optimization, we first simulate them using the GME method at a higher resolution ($g_{\text{max}} = 3$, aggregating losses over a $3 \times 3$ array of reciprocal lattice vectors $\vec{k}$). We plot the resulting far-field radiation patterns in the top panels of Figures \ref{fig:GME_FDTD_comparison}a--h, with the quality factor and coupling efficiency indicated in the insets. While the quality factor for all designs remains $> 8.7 \times 10^4$, the coupling efficiency falls between 0.67 and 0.75 for all designs, indicating a limitation in achieving high coupling efficiency with this GME resolution. We note that improving this performance further would require a higher resolution and the incorporation of higher-order guided modes during the optimization, which incurs significantly higher computational resource costs. Specifically, the memory requirement scales as $\mathcal{O}(g^4_{\text{max}})$ and the simulation time scales approximately as $\mathcal{O}(g^6_{\text{max}})$ \cite{minkovInverseDesignPhotonic2020, zanottiLegumeFreeImplementation2024}.}

\rev{We validate the final designs by performing FDTD simulations and plotting the obtained far-field radiation patterns in the bottom panels of Figures 9a--h. The quality factor across all NA$_{t}$ values ranges between $4.4 \times 10^4$ and $6.6\times 10^4$, which is up to a factor of two smaller than the GME results and shows no correlation with NA$_t$. This reduction in the quality factor is expected and is attributed to the approximate nature of the GME method \cite{minkovInverseDesignPhotonic2020, zanottiLegumeFreeImplementation2024}.}

\rev{We also observe that designs with NA$_{t} \le 0.5$ exhibit a Gaussian-like radiation pattern with high coupling efficiency, closely matching the GME far-field radiation pattern. However, for NA$_{t} > 0.5$, the far-field radiation pattern differs significantly from the GME results. This discrepancy arises because resolving the Gaussian far-field at higher target NAs requires smaller feature sizes that fall below the effective resolution limit of the current GME simulation. Consequently, achieving high coupling efficiencies for NA$_{t} > 0.5$ would require a higher GME resolution ($g_{\text{max}} > 3$) during the inverse design optimization, which becomes computationally prohibitive.}

\rev{\subsection{\label{app_subsec:physical_mech} Physical mechanism behind inverse design}}

\rev{A high quality factor of the photonic crystal cavity implies a lower total amount of out-of-plane radiation. However, this does not necessarily result in a low coupling efficiency. Our inverse design optimization framework does not aim to increase the total amount of out-of-plane radiation; rather, it modifies the L3 cavity to shape the outgoing radiation in the far field. By tailoring the far-field to a specific NA, we ensure that the emitted radiation couples efficiently into the collection optics. To demonstrate this, we plot the amplitude of the Fourier transform spectra of the initial L3 and inverse-designed cavity modes, obtained using FDTD, in Figures 10a and 10b, respectively. The white circles in the two panels represent the light cone. The Fourier transform spectrum of the L3 cavity contains large components at diverging angles, resulting in a lower quality factor and poor coupling efficiency. On the other hand, the Fourier spectrum of the inverse-designed cavity features smaller components concentrated at lower angles, simultaneously yielding a high quality factor and a Gaussian-like far-field radiation pattern.}

\begin{figure}[hbt]
    \centering
    \includegraphics[width=\linewidth]{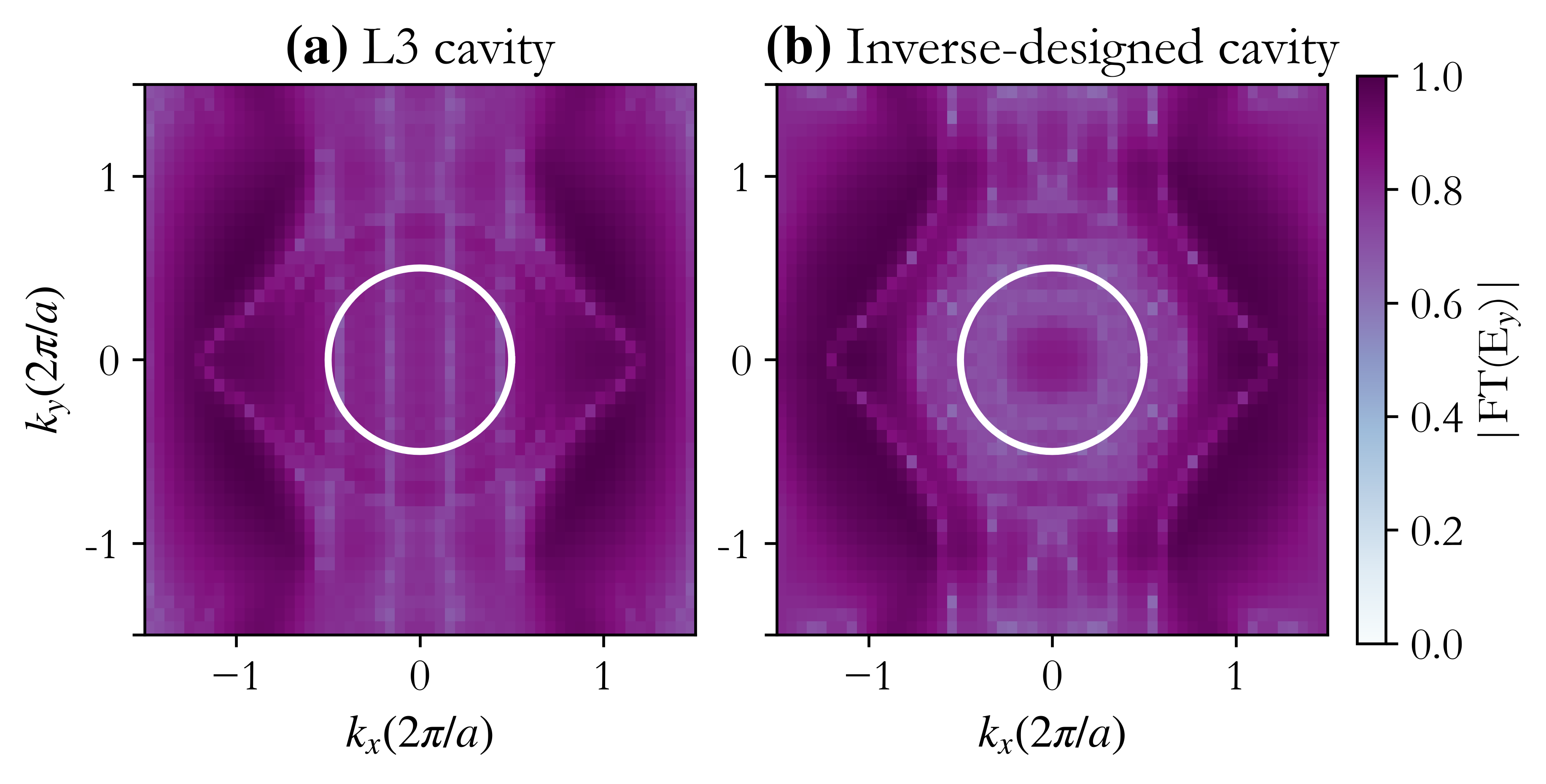}
    \caption{\rev{\textbf{Fourier transform of the cavity mode.} Panels (a) and (b) represent the amplitude of Fourier transforms of the FDTD-simulated cavity modes for an initial L3 and an inverse-designed photonic crystal cavity, respectively. The white circles indicate the light cone.}}
    \label{fig:FT_comparison}
\end{figure}

\rev{\section{\label{app_sec:sidewall angle}Effect of imperfect sidewall angle on cavity performance}}

\rev{A sidewall angle $\neq 90^\circ$ breaks the mirror symmetry of the photonic crystal along the slab thickness direction. This symmetry breaking mixes the TE and TM cavity modes and results in additional radiative loss. This effect further reduces the performance of the photonic crystal cavity.}

\begin{figure}[hbt]
    \centering
    \includegraphics[width=\linewidth]{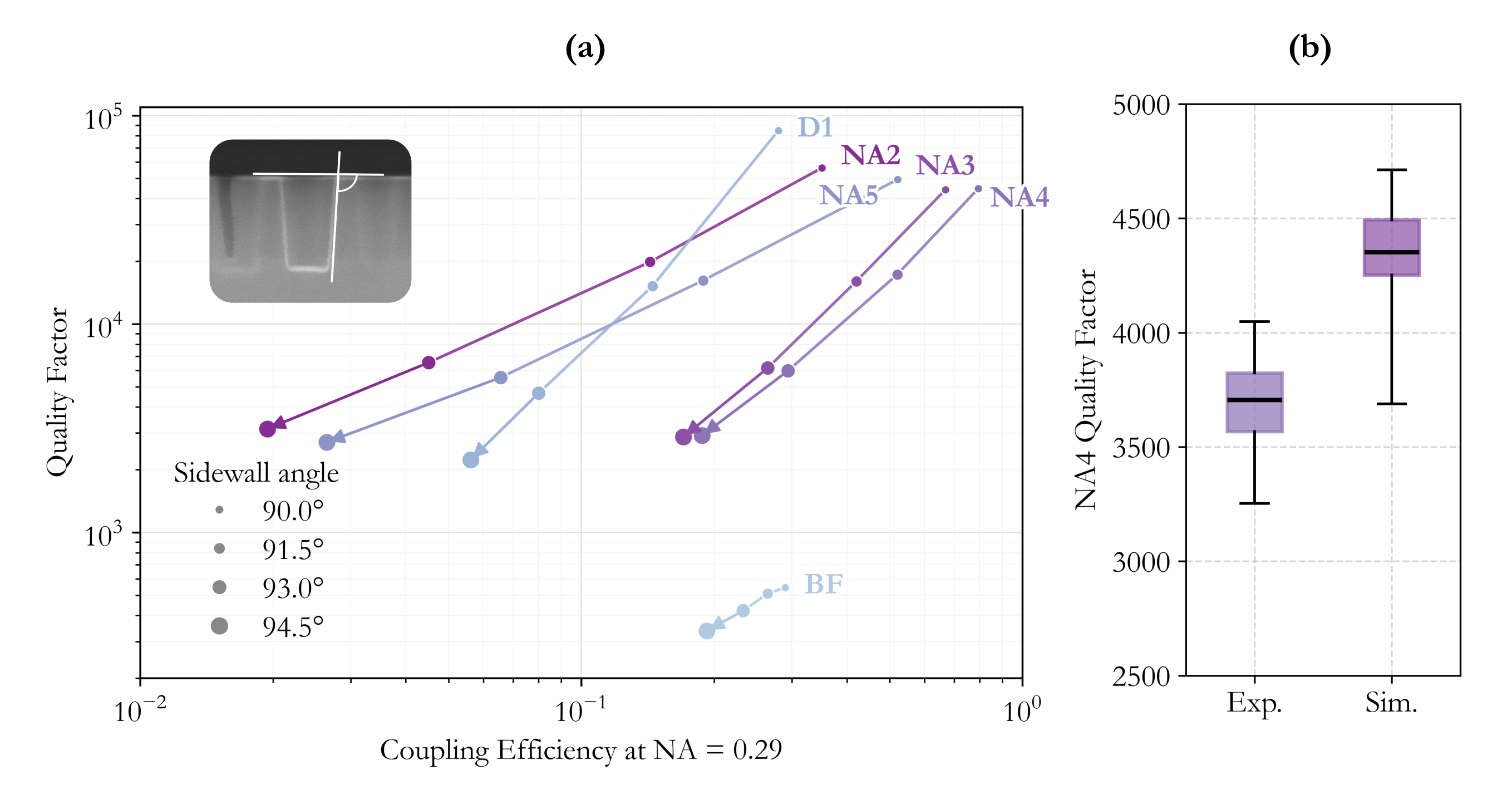}
    \caption{\rev{\textbf{Effect of tilted hole sidewalls on cavity performance.} (a) Simulated quality factor and coupling efficiency for various cavity designs as a function of sidewall angle (ranging from $90^\circ$ to $94.5^\circ$, indicated by increasing marker size and arrow direction). The performance of the inverse-designed cavities degrades significantly, whereas the band-folding (BF) design remains relatively robust. Inset: SEM image of a cleaved hexagonal hole array used to estimate the experimental sidewall angle, revealing an average angle of $93.5^\circ$. (b) Comparison of the experimentally measured quality factor for the NA4 design with the simulated theoretical model (Eq. \ref{eq:noisy_Q}) that accounts for the ideal design, hole radius disorder, and tilted sidewalls.}}
    \label{fig:sidewall_angle}
\end{figure}

\rev{To estimate the sidewall angle in our measured devices, we fabricate a new sample with a long hexagonally arranged hole array of the same diameter (124.8 nm) as the photonic crystal cavity to replicate the precise nanofabrication conditions of the measured sample. We cleave the sample to expose the etched hole cross-section and image it using an SEM (Figure \ref{fig:sidewall_angle}a inset). The SEM images reveal sidewall angles ranging from $93^\circ$ to $95^\circ$, with an average sidewall angle of $93.5^\circ$. We note that the sidewall angle of the etched holes can be further reduced in the future by optimizing the nanofabrication recipe.}

\rev{Figure \ref{fig:sidewall_angle}a illustrates the effect of these tilted sidewalls on the performance of the photonic crystal cavity designs. We plot the quality factor and coupling efficiency at NA = 0.29 of the various cavity designs as a function of an increasing sidewall angle, ranging from $90^\circ$ to $94.5^\circ$. The performance of the inverse-designed cavities degrades significantly as the sidewall angle increases (indicated by the increasing marker size and arrow direction). However, the BF design remains robust, with its performance decreasing at a much slower rate than the inverse-designed cavities. Specifically, for the NA4 cavity design, the quality factor drops from $4.5 \times 10^4$ to $5 \times 10^3$ at a sidewall angle of $93^\circ$. By treating each imperfection as an independent loss channel, the combined effect of hole radius disorder and tilted sidewalls on the cavity quality factor can be accounted for by the following formula \cite{minkovStatisticsDisorderinducedLosses2013}:}

\begin{equation}
    \label{eq:noisy_Q}
    \rev{\frac{1}{Q_{\text{Sim.}}} = \frac{1}{Q_{\text{design}}} + \frac{1}{Q_{\text{disorder}}} + \frac{1}{Q_{\text{sidewall}}}}
\end{equation}

\noindent \rev{where the reciprocal $Q$ values correspond to the loss rates of the cavity mode along different channels. The value of $1/Q_{\text{design}}$ is set by the ideal photonic crystal cavity design, $1/Q_{\text{disorder}}$ arises from the disorder in the hole radius, and $1/Q_{\text{sidewall}}$ represents the contribution from tilted hole sidewalls. In Figure \ref{fig:sidewall_angle}b, we plot the simulated noisy quality factor $Q_{\text{Sim.}}$ alongside the experimentally measured quality factor for the cavity design NA4. The median quality factor for the experimental data (Q = $3.7 \times10^3$) matches well with the simulated noisy quality factor (Q = $4.3 \times10^3$), demonstrating that our disorder model accurately replicates the experimental results.}

\rev{The coupling efficiency of the inverse-designed cavities at NA = 0.29 also displays a sharp decrease with increasing sidewall angle, with the exception of the BF cavity design. In particular, for a sidewall angle of $94.5^\circ$, the coupling efficiency of the BF design drops from 0.29 to 0.19, whereas the NA4 cavity design experiences a much steeper drop from 0.66 to 0.16. This robustness qualitatively explains the high coupling efficiency measured experimentally for the BF design. These results isolate the effect of imperfect sidewall angle on the coupling efficiency alone. As there is no simple analytical formula for the combined effect of hole-radius disorder and sidewall tilt, a fully quantitative treatment would require additional simulations. Such simulations are computationally prohibitive for two reasons: capturing disorder statistically requires ensemble-averaging over many random realizations, and the disorder breaks the mirror symmetries along the $x$, $y$, and $z$ axes, which are present in the ideal design, thus substantially increasing the cost of each individual simulation.}
\bibliography{references}
\bibliographystyle{IEEEtran}
\end{document}